\documentclass{WileyMSP-template}
\usepackage[utf8]{inputenc}
\usepackage{amsmath}
\usepackage{amsfonts}
\usepackage{amssymb}
\usepackage{graphicx}
\usepackage{algorithm,algpseudocode}
\usepackage{listings}
\usepackage{subcaption}
\usepackage{hyperref}
\usepackage{xparse}
\usepackage[style=numeric, sorting=none, maxnames=6, minnames=3, backend=biber, giveninits=true]{biblatex}
\usepackage{fixltx2e}
\usepackage{doi}

\newif\ifnote

\newcommand{\figref}[1]{\textbf{Figure \ref{#1}}}
\notetrue

\begin{document}
\title{Multiscale Microscopy via Automation: Dual Magnification ESEM Imaging by Frame Alternation}
\maketitle

%{Maurits Vuijk$^1$ \and Johannes Zeininger$^{1, 2}$ \and Luis Sandoval$^1$ \and G\"unther Rupprechter$^2$ \and Beatriz Roldan Cuenya$^1$ \and Karsten Reuter$^1$ \and Thomas Lunkenbein$^{1,3}$ \and Christoph Scheurer$^{1,4}$ \\
\author{Maurits Vuijk}
\author{Johannes Zeininger}
\author{Luis Sandoval}
\author{G\"unther Rupprechter}
\author{Beatriz Roldan Cuenya}
\author{Karsten Reuter}
\author{Thomas Lunkenbein*}
\author{Christoph Scheurer*}

\begin{affiliations}
Maurits Vuijk, Dr. Christoph Scheurer, Prof. Dr. Karsten Reuter\\
Theory Department, Fritz-Haber-Institut der Max-Planck-Gesellschaft, Faradayweg 4-6, 14195 Berlin, Germany

Dr. Johannes Zeininger, Prof. Dr. Beatriz Roldan Cuenya \\
Interface Science Department, Fritz-Haber-Institut der Max-Planck-Gesellschaft, Faradayweg 4-6, 14195 Berlin, Germany

Maurits Vuijk, Dr. Luis Sandoval, Prof. Dr. Thomas Lunkenbein\\
Inorganic Chemistry Department, Fritz-Haber-Institut der Max-Planck-Gesellschaft, Faradayweg 4-6, 14195 Berlin, Germany

Prof. Dr. G\"unther Rupprechter, Dr. Johannes Zeininger\\
Institute of Materials Chemistry, TU Wien, Getreidemarkt 9, 1060, Vienna, Austria

Prof. Dr. Thomas Lunkenbein\\
Chair of Operando Analytics for Electrochemical Energy Storage, University of Bayreuth, Universitätsstraße 30, 95447 Bayreuth, Germany\\
Email: thomas.lunkenbein@uni-bayreuth.de

Dr. Christoph Scheurer\\
Institute of Energy Technologies, Grundlagen der Elektrochemie (IET-1), Forschungszentrum J\"ulich GmbH, J\"ulich, Germany\\
Email: scheurer@fhi.mpg.de
\end{affiliations}

\date{July 2025}

%\corres{\name{Christoph Scheurer} \email{scheurer@fhi.mpg.de}}
%\corres{\name{e} \email{e}}


%\selfcitation{}

%\received{xx xxxx xxxx}
%\revised{xx xxxx xxxx}
%\accepted{xx xxxx xxxx}
\begin{abstract}
In Environmental Scanning Electron Microscopy (ESEM) experiments, the acquisition parameters are generally kept constant throughout the collection of a data set. This limits data collection to one data set at a time, and frequent human interaction is required to maintain the image quality. Here, we use a custom-designed automation interface to minimize such supervision and allow for the collection of multiple interlaced data sets simultaneously. The oscillatory modes of an example catalytic system (hydrogen oxidation over Co foil) were employed as a tunable spatiotemporal test case. Using our automation interface, we can implement more advanced acquisition programs into the microscope that allow dual imaging --- effectively bridging reaction monitoring between different length scales. By using automation to change the settings of the acquisition after each frame, we are able to capture alternating magnifications of the same process and sample location. Both a low magnification overview of the mesoscopic surface dynamics and a high magnification field of view of the ongoing structural changes of a selected surface motif were acquired. Using such truly correlative data captured with the dual magnification method, cross-scale correlations about catalytic systems including phase information can be established.
\end{abstract}

\section{Introduction}
%\mynote{1) What is the problem we want to address?}
Detailed knowledge of catalyst surfaces and their behaviour under active conditions is essential in uncovering their underlying dynamics and functionality. \cite{Chee2023} Particularly the investigation of active phase transitions and areas around prominent dynamic structural features and defects are of special interest when it comes to investigating catalytic activity. \cite{SandovalDiaz2024} \cite{Suchorski2018} Such points of interest are typically scattered across the catalytic system, or either present or visible only at certain times, and occupy a range of sizes within the same system. \cite{Vogt2022} \cite{Zeininger2021} The situation is complicated by the fact that catalysis is a multidimensional scientific discipline and technology that spans several size and length scales. For a complete understanding, information from at least six dimensions must be combined and correlated: the x-, y- and z-directions, the time domain, the length scale, and the function. It is important to observe them in parallel as these processes might by related to synergistic effects\cite{Schloegl2015} being part of the, so called complexity gap \cite{Hansen2023} \cite{Chee2023}. Many microscopy techniques are limited in their ability to find and focus on such spatiotemporally rare features, let alone monitor multiple scales over long periods of time. The volume of data collected can also cause issues in identifying the relevant captured frames in the dataset. \cite{Alcorn2023} To investigate nontrivial multiscale dynamic behavior, certain processes must be locally resolved and differentiated from collective monitoring. 
To this end, Environmental Scanning Electron Microscopy (ESEM) is an important technique when looking at dynamics of active catalysts, allowing imaging of surfaces evolving under reaction conditions. Using ESEM, resolutions in the ~10nm regime can be achieved under optimal conditions, but lower magnifications are also commonly used. The principle of ESEM is similar to regular SEM, but a series of pumps create a pressure gradient in the column, allowing for regular high vacuum operation of the gun despite gas flow in the chamber. \cite{Danilatos1993} How can we use the ESEM for multiscale imaging?

\begin{figure*}[th]
    \centering
    \includegraphics[width=\textwidth]{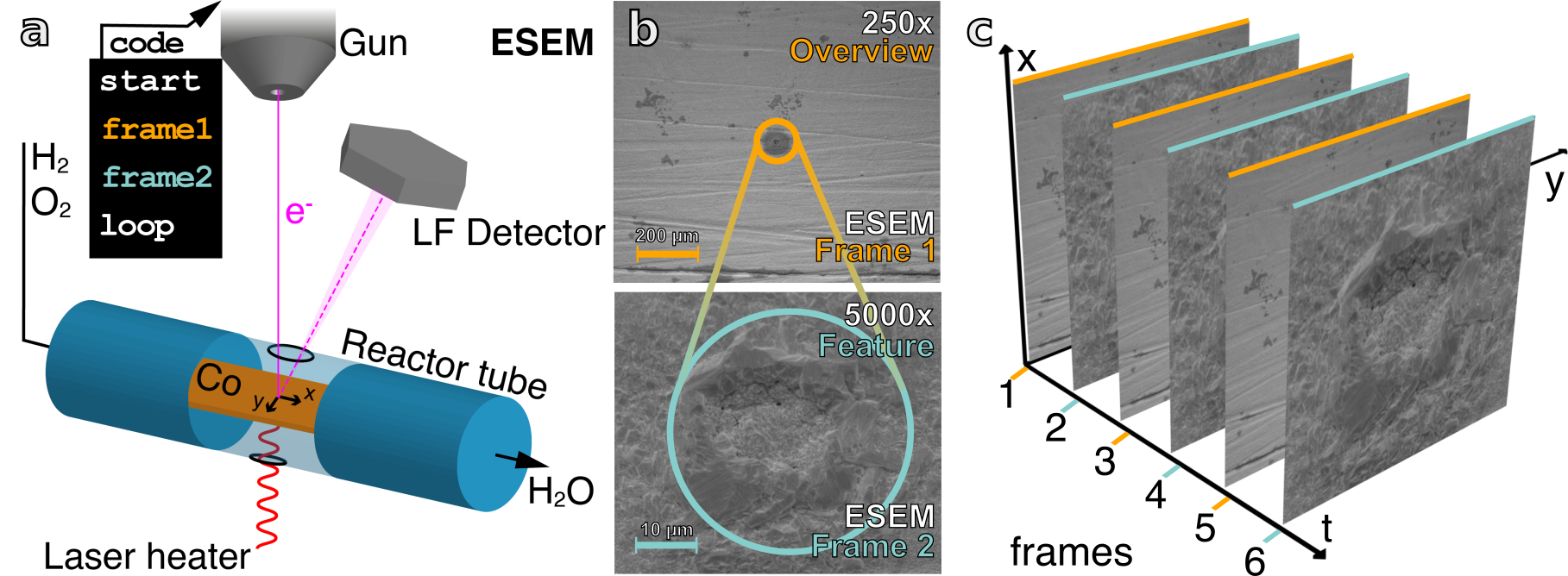}
    \caption{\textbf{a)} ESEM setup. The Co sample is placed in the reactor on a movable stage under H$_2$ and O$_2$ gas flow and laser heating. Using the large-field detector of the ESEM, images are captured. By using an automation interface, the magnification is changed after each captured frame. \textbf{b)} Two different regimes of the observed sample captured by ESEM. \textbf{Overview:} On the surface around the central feature, oscillating features appear and fade out on the surface. \textbf{Feature:} The feature (a magnified field of view of the central area of the \textbf{Overview} image) undergoes development and oscillating surface effects. The goal is to acquire fields of view of both of these regimes simultaneously. \textbf{c)} By alternating the magnification as described, the resulting captured data consists of two interleaved sequences capturing a field of view of the sample, each with a different magnification.}
    \label{fig:grabs}
\end{figure*}

Using the ESEM, it is possible to collect time series image data of catalyst surface dynamics under active conditions. When collecting such data in ESEM experiments, the common procedure is that the operator explores the surface to find a view of a specific feature on the sample. The image quality is then optimized (magnification, focus, and other imaging settings). Then, images are captured continuously for the duration of the experiment with human intervention as needed. \cite{Wang2015} \cite{Barroo2020}  The result of conducting an experiment in this fashion is a data set capturing the local dynamics of the system, ideally with constant settings (magnification, position, etc). In the case of dynamics with a characteristic timescale much slower than the scan time of the microscope, this results in redundant information being recorded. Although such data sets give insights into the dynamics surrounding the feature, it may result in a myopic view of the sample: vital information obtained from a different field of view (magnification or location) may be missed. To obtain additional data, the experiment can be extended or repeated with a different field of view, however statistics over different experiments will be harder to correlate (e.g. with external stimuli) due to the added dimensions in the data. In particular, multi-resolution auto- and cross-correlation information about the time-series data can not be retrieved in this often practiced mode of experimentation. Furthermore, due to the image drift induced by temperature changes, it can be difficult to keep a feature centered in the field of view. \cite{Podor2019} We propose a method that allows for the simultaneous collection of multiple interleaved data sets of the same sample with different capture parameters, through alternation of the relevant imaging parameters (in this case the magnification). Using this method, multiple scales can be probed in a single experiment, allowing the direct correlation of the evolution of a feature between different scales. Furthermore, as the amount of captured frames remains the same, data management is not an issue. A schematic overview of the method is given in \figref{fig:grabs}.

%\mynote{2) What has been done in literature before, and HOW and What they found?}
The use of automation interfaces for scripting to perform more complex acquisition routines has been employed in the field of electron microscopy for static imaging modes before. Using the SerialEM software \cite{Mastronarde2005} as an interface to the TEM, custom Python functions allow for automatic gathering of images at series of user-defined target points.  \cite{Schorb2019} It has been shown that using custom programmed scan patterns can reduce the damage caused by beam exposure in TEM experiments. This strategy uses an automation interface to the TEM to guide the acquisition beam in a user-defined pattern \cite{Velazco2022}. An outlook by \textit{Kalinin et al.}\cite{Kalinin2023} suggests a need for real-time machine learned control of electron microscopy systems. To this end, they suggest solutions to challenges related to data storage, computing power and hardware interfacing that may arise when implementing automated microscopy systems. To the end of a modern programming interface, ImSwitch \cite{CasasMoreno2023} aims to provide a unified interface for microscopes in biochemistry. Included in this system is a scripting interface capable of executing sequential stage movements and capture operations. The AutoEM system \cite{osti_2221791} represents the state-of-the-art automated TEM system: Using an automation interface, they are able to employ computer vision models to recognize particles within an initial systematic scan. Then, in subsequent operation of the TEM, the AutoEM system is able to apply the same computer vision models (retrained with few-shot learning to recognize specified particles) to drive the field of view towards the selected particles of interest. 
In the present work, a programming interface similar to the one of ImSwitch is described. The primary challenge is the dynamic system being observed requiring more care to integrate the state of the ongoing experiment into the code. We have previously investigated the use of computer vision models, which could be applied to this problem. \cite{Vuijk2025}

There exists little previous literature into automated, smart control of the microscope during experiments involving dynamic sampling (methods involving multiple frames over time of a changing surface). Most research in the area of "automated microscopy" focuses on optimization of the image and subsequent analysis of the image. This lack of research has several likely reasons. First, there is a lack of vendor support for microscopy automation - the inclusion of automation interfaces into microscopes has been a recent change. Furthermore, such packages do not tend to come with the microscope, and are additional purchases, decreasing their availability.  Also, errors in the code may lead to costly damage to the hardware. \cite{Kalinin2021} Lastly, there is lesser benefit gained from automation in the case of static sampling (methods involving single frames or multiple frames of a non-changing surface) as observed in many microscopy techniques. The \textit{in-situ} nature of ESEM allows for real-time automation of the dynamic sampling. With advances in the availability of automation interfaces on ESEM devices, we hope to see more research in the field of automated dynamic sampling in (electron) microscopy. \cite{Vesseur2018}

%\mynote{3) What are the remaining issues, and what is still missing?}

The main challenge is managing the automation of a dynamic system. As such, this article presents our initial insights into this area of research, using an algorithmic method of automation. We plan on researching this topic further, introducing more complex algorithms for microscope automation. The goal is to encourage the use of automation interfaces for microscopy, furthering contributions to the area of automated observation of dynamic systems in microscopy.
%\mynote{4) What do you propose?}

In our approach, we make efficient use of all the capture time of the microscope by using an automation interface to alternate the settings of the image between capturing frames. Doing so, it is possible to capture the surface dynamics of the sample at multiple magnifications, positions, voltages, or other combinations of such settings. In this way, it is possible to obtain high resolution images of a feature as well as the evolution of the surface around it, monitor multiple features, and thereby obtain additional insight. 

%\mynote{5) What have you done, and what have you found with this new approach?}
In this work, this technique was used to capture the oscillatory dynamics of a Co catalyst surface at 550 \textdegree C under flow of hydrogen and oxygen at multiple magnifications. The two fields of view provided by the dual magnification technique give insight into the overall surface dynamics and feature dynamics during the oscillating oxidation and reduction of the catalyst surface.

\section{Methods}
The system we have investigated using ESEM is a polycrystalline Co foil catalyst surface under a constant gas flow of a mixture of 50:1 hydrogen to oxygen at high vacuum pressure ($10^{-3}$ Pa). Using a laser heating setup, the sample was heated to 550 \textdegree C and kept at a constant temperature. At this temperature, self-sustained oscillations are visible on the surface. \cite{Saraev2015}
A custom automation interface to the FEI Quanta 200FEG ESEM was developed, allowing us to control by code what is normally possible through the graphical user interface. On modern devices, it is generally possible to use a vendor-supplied automation interface (such as ThermoFisher AutoScript \cite{geurts2024leveraging}) for this purpose. Using the automation interface, we can then use a common programming language (in this case Python \cite{10.5555/1593511}) to control the microscope.

\begin{figure}    
\centering
    \includegraphics[width=0.5\textwidth]{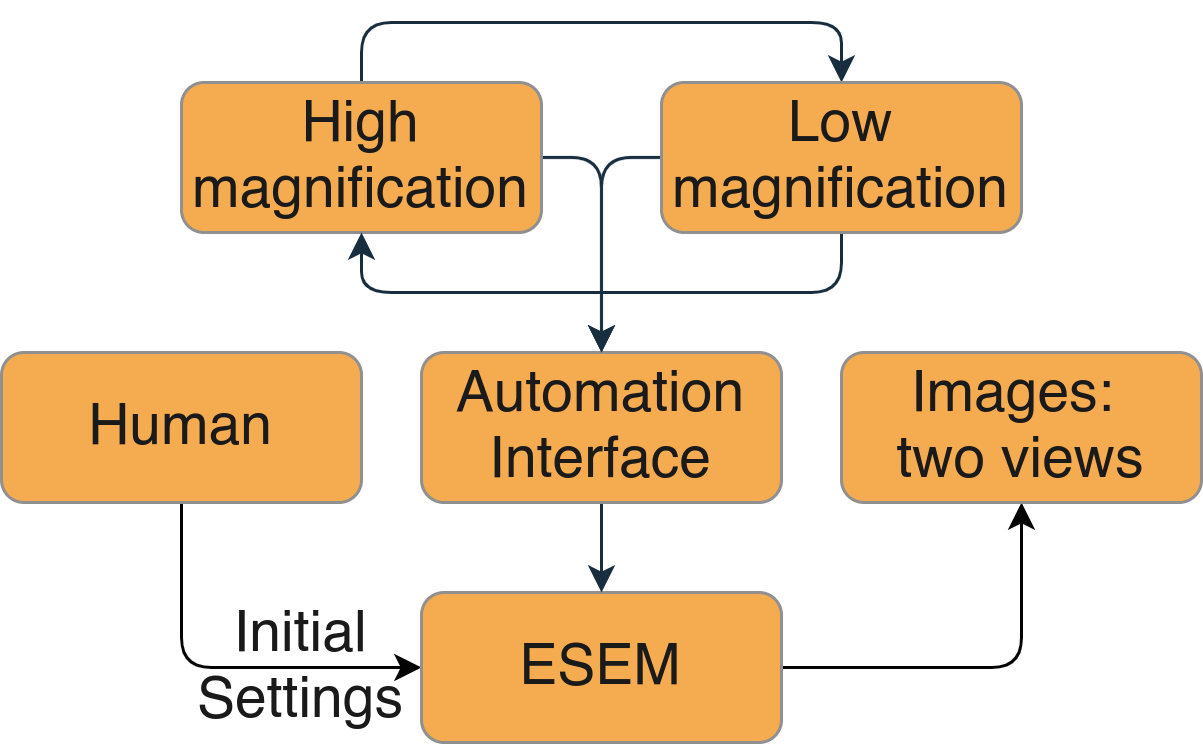}
    \caption{Schematic showing the implementation of the dual magnification workflow on the ESEM. In addition to alternating magnification, other ESEM parameters could be automated, resulting in other insights. Changing the voltage would provide additional data with regards to the surface material properties and topography. Changing the stage position would provide additional data in the form of other points of view of the sample. Changing the working distance would allow for 3D reconstruction of the surface geometry.}
    \label{fig:diag2}
\end{figure}

In order to acquire images with alternating magnification, we use the automation interface to set the magnification to 250x, then instruct the microscope to take a frame. After the microscope has collected this frame, the magnification is set to 2500x or 5000x (different high magnification values were used during the experiment). These magnifications were selected in accordance with the monitored reaction dynamics. Following this, a delay is introduced, as the microscope needs some time to refocus after the magnification is changed. The defocus is negligible in the low magnification value chosen in this case, thus the delay is only present after changing to high magnification. Then, another frame is taken at the high magnification. This cycle is repeated, capturing alternating low-magnification and high-magnification frames until interrupted manually. With the scan time of 40 seconds per frame, this results in a 120-second cycle, where each cycle gives a low magnification and a high magnification frame. Pseudocode for this sequence is given in \textbf{Algorithm \ref{alg:alg1}}, and a schematic for this procedure is given in \figref{fig:diag2}.
\begin{algorithm}
\caption{Image Capture Loop}
\begin{algorithmic}
\While{true}
    \State \textit{SetMagnification}(Low Magnification)
    \State \textit{CaptureImage}()
    \State \textit{SetMagnification}(High Magnification)
    \State \textit{Delay(T)}
    \State \textit{CaptureImage}()
\EndWhile
\end{algorithmic}
\label{alg:alg1}
\end{algorithm}

This approach describes only one possible use for algorithmic automation on the ESEM. It would also be possible to take a frame at a low magnification as an overview, then change to a high magnification and take three frames at different stage positions of different features. Another approach could change the voltage between frames. This way, different depth sensitivities could be gathered simultaneously. This approach allows for the gathering of large amounts of data during a single experiment, reducing the need for repeated measurements and improving the consistency of the gathered data. This could be implemented directly using Python using a similar loop to \textbf{Algorithm \ref{alg:alg1}}, with more steps. For ease of use by non-programmers, a system where the desired sequence is described using a graphical user interface could be implemented.

\section{Example}

\begin{figure*}
    \centering
    \includegraphics[width=\textwidth]{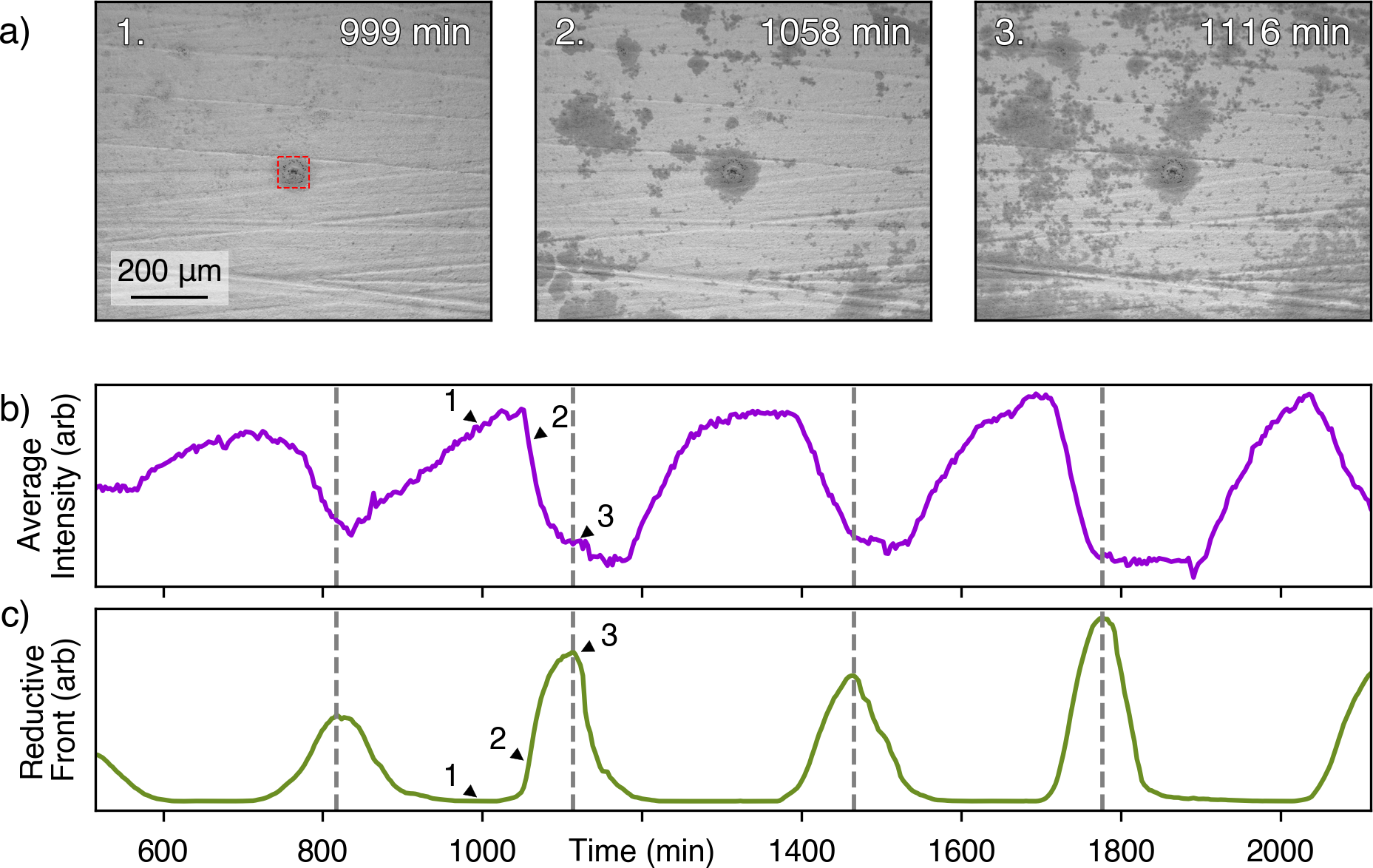}
    \caption{Low magnification results. \textbf{a)} Surface images showing the fully oxidized surface (bright regions) in \textbf{1}, with high magnification feature outlined in red, initial reductive areas (darker regions) in \textbf{2} and the maximally reduced state  in \textbf{3}. \textbf{b)} Overall image intensity per frame taken as the average intensity of all pixels of each frame.  \textbf{c)} Reductive front showing the presence of newly reduced areas. Calculated as the area of the negative component of a background subtraction of a previous frame, capturing the darkened (reduced) areas. This does not include the re-oxidation, which is present as the subsequent increase in \textbf{b}.}
    \label{fig:lmr}
\end{figure*}

\begin{figure*}
    \centering
    \includegraphics[width=\textwidth]{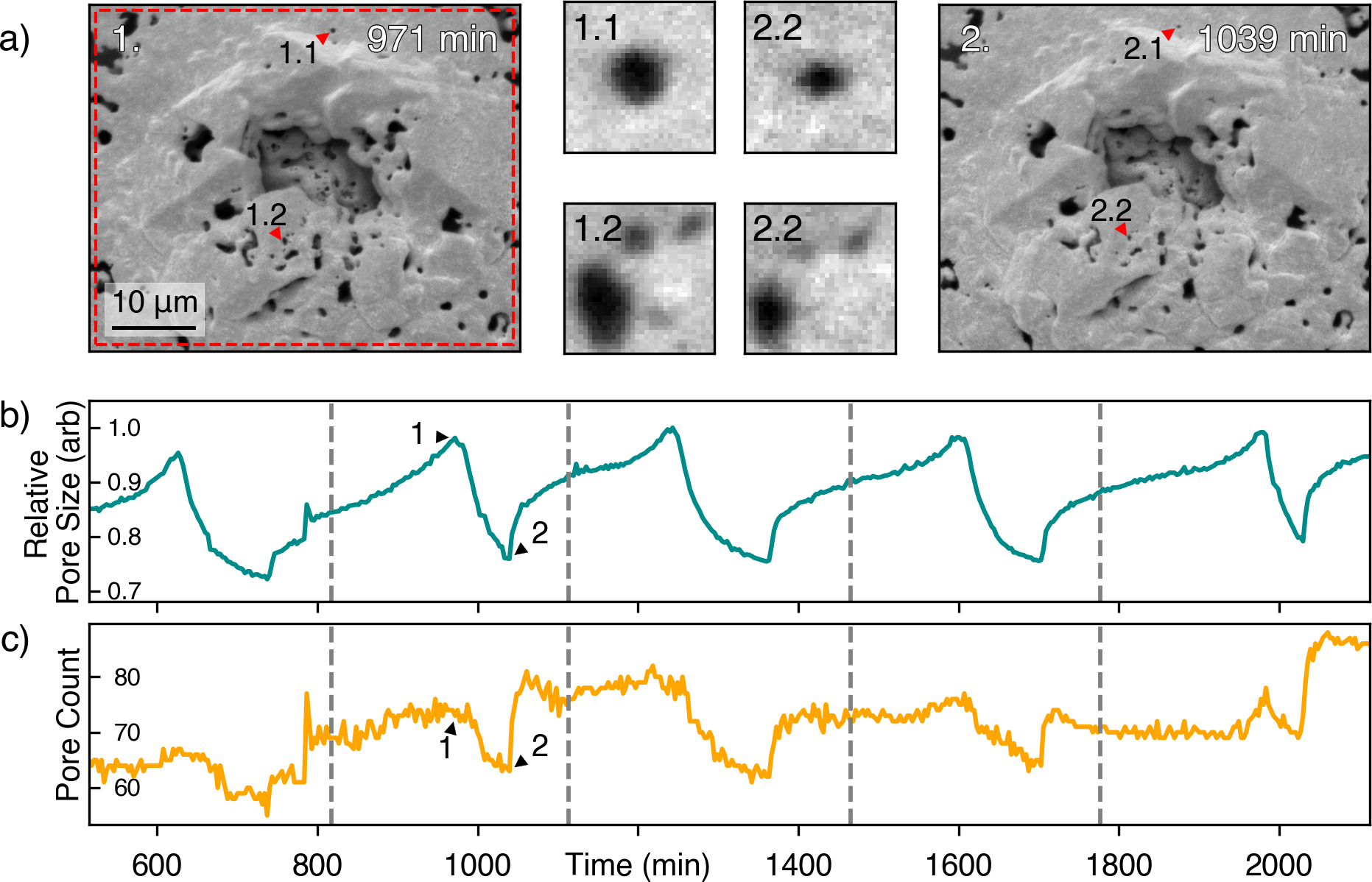}
    \caption{High magnification results. \textbf{a)} Frame \textbf{1} at the largest relative surface visible pore size and frame \textbf{2} at the smallest relative surface visible pore size. To illustrate the changing pores, digitally magnified regions of interest with changing pores are given. \textbf{b)} Quantification of the surface visible pore area. The pores were segmented using a difference of Gaussians approach of selected areas of the image, the scale is normalized to the highest pore area coverage. \textbf{c)} Count of individual visible pores visible on the surface.}
    \label{fig:hmr}
\end{figure*}

Using the dual magnification program on the microscope, we were able to collect two interlaced data sets of the same system effectively simultaneously. This allows for a truly correlative analysis of the surface, including phase information across the two captured magnifications. \cite{Zeininger2022} \cite{Winkler2023} Modes of oscillatory behaviour on the polycrystalline cobalt foil surface under 50:1 H$_2$:O$_2$ were explored using the ESEM. As we were able to tune the timescale of the dynamics to match the scan time of the microscope, the time-resolution remained sufficient to capture the full extent of the surface dynamics despite the effective threefold reduction in scan rate per captured sequence. As such, we gain insights in the global catalytic behaviour, which is reliant on processes at different scales, by capturing truly correlative and complementary monitoring viewpoints.
With our method, we were able to capture dynamics and structural evolution of an extended surface area (\figref{fig:lmr}) and a selected specific feature at the catalyst surface,  a protrusion with a group of pore openings (\figref{fig:hmr}).

The progression of the mm-scale surface reduction can be seen in \textbf{\ref{fig:lmr} a)}: From the oxidized surface (corresponding to the regions of high image intensity) (\textbf{1}), reduced areas (of lower image intensity) appear in many places across the surface (\textbf{2}), which then expand radially outwards until the expansion stops (\textbf{3}). The image intensity was linked to the surface state by previous gas flow experiments (see \textbf{SI 1}), allowing us to use image processing techniques to translate image contrast to surface state information. The intensity over time (calculated as an average of each frame of the low magnification series) and reductive front (calculated as a background subtraction of the low magnification series where only negative results are kept, thereby quantifying the expansion of the reductive front.) in \textbf{\ref{fig:lmr} b), c)} show that these processes occur as part of an oscillating cycle. In \textbf{SI 2} \cite{3.YFJ9MR_2025}, the full dual magnification dataset as two correlated videos of the oscillatory processes are given. The reduced areas exhibit synchronous behaviour in that there is a certain time period in which the process takes place. However, there is also an asynchronous layer exhibiting anisotropic and spatiotemporal behaviour: within the reductive part of the cycle the initial appearance of each reduced area differs. Following the reductive part of the cycle, there is a fully synchronous oxidation of all the reduced areas. The reduced areas appearing on the surface are not randomly distributed, they instead reappear and fade out in similar locations during each cycle. There are large areas of the surface (visible in \textbf{SI 2}) that do not participate in the reduction/oxidation cycle. Conversely, the location of reductive areas originate from defects, such as previously present pores and surface ridges. 
While all of the reductive areas are part of the same overall cycle, there is a degree of asynchronicity in the initial appearance: the reductive areas nucleate on the surface at different times within the reductive time window. Conversely, the following surface re-oxidation occurs as a synchronized oxidation of all reduced areas on the surface. 

A high magnification view of the central feature of the low magnification image is visible in \textbf{\ref{fig:hmr} a)}. A large amount of macropores are visible on the surface. As a part of the oscillatory cycle of the surface, these macropores exhibit growth and shrinkage. The total area of the visible macropores in the high magnification image was quantified per frame by segmenting the pores using a difference of Gaussians approach and is visible in \textbf{\ref{fig:hmr} b)}. It can be seen that there exists a short time window of pore shrinkage and a long time window of pore growth in each cycle. During the pore shrinkage the larger pores shrink in diameter and the smaller pores close entirely, as seen in \textbf{\ref{fig:hmr} c)}, reappearing in different positions during the next growth time window. From the maximum (\textbf{1}) to the minimum (\textbf{2}) average pore size the cross section area of the pores is about 27\% smaller, which may not be immediately obvious when comparing \textbf{\ref{fig:hmr} a) 1 and 2}. From comparing the digitally zoomed-in \textbf{a) 1.1, 1.2} with \textbf{a) 2.1, 2.2}, the difference becomes more visible. The observed shrinkage percentage is below the maximum geometrical unit cell expansion of Co to oxide. \cite{Jain2013}

\begin{figure*}
    \centering
    \includegraphics[width=\textwidth]{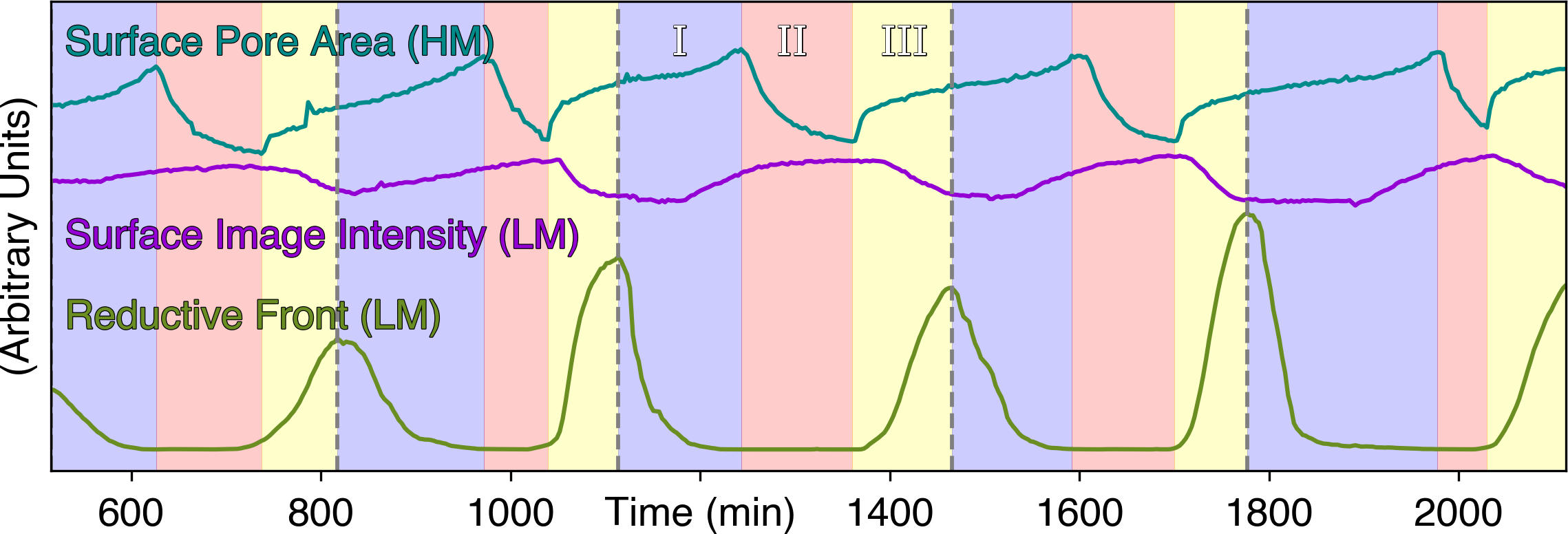}
    \caption{Multiscale correlation of dual magnification timeseries data. The surface pore area quantifies the total surface visible pore area in each frame. The surface image intensity serves as a reference for the proportion of reduced area in the frame. The reductive front quantifies the newly reduced areas. The re-oxidation is not visible in this line. The oscillatory cycle is divided into three sections by the beginning and end of the pore shrinkage, and the peak of the reductive front, which indicates the end of the reduction. The sections are: \textbf{I:} Pore growth and re-oxidation. \textbf{II:} Pore shrinkage and re-oxidation. \textbf{III:} Pore growth and reduction. }
    \label{fig:rgraph}
\end{figure*}

In \figref{fig:rgraph}, a comparison between the data obtained from both the high and low magnification data is visible. By using the dual magnification method, the phase shift between high and low magnification events is preserved, allowing direct comparisons between the datasets. The low magnification dataset provides an indicator of the current surface state by the localized reductive front and reduced area, and the high magnification dataset provides a proxy indicator of the current pore system by the surface pore area. As a result of the phase difference between the oscillations of the pore area and the surface reduction, three sections are present based on the current direction of the respective graphs: \textbf{I:} pore growth and re-oxidation, \textbf{II:} pore shrinkage and re-oxidation, and \textbf{III:} pore growth and reduction. The combination of pore shrinkage and reduction was not observed. Pore shrinkage and re-oxidation, and pore growth and reduction are both expected, as this change in surface phase directly affects the pore size due to the differences in density between the oxide and the reduced metal. The unexpected combination, then, is pore growth and re-oxidation (\textbf{I)}. The timescale of this period (over 120 minutes) is much larger than the experimental time resolution of the dual magnification method (2 minutes), as such this observation is not an artifact introduced by the time delay during the frame acquisition.

Due to this divergence of pore size and surface oxidation state, we conclude that a compartment other than only the 2-dimensional surface must be involved in the dynamic processes. An additional dimension into the depth of the sample, the pores, provide additional reaction compartments that give rise to the decoupling of reaction behaviours. This could explain the unexpected combination of observed trends in \textbf{I)}. Following the reductive front propagation events in \textbf{III}, we observe a gradient in redox behaviour in the new cycle in \textbf{I}. The surface is already oxidizing while the pore openings connected to the pore network are still reducing. This hydrogen buffer below the surface slowly depletes during the cycle, affecting regions of the catalyst more strongly depending on their distances into the pores. This effect resembles diminishing hydrogen spillover. This sequence is followed by full oxidation in \textbf{II}, until the next transition to a reduced surface state in \textbf{III} abruptly occurs. This nucleation of reductive fronts below the oxidised surface appears on the ESEM images similar to the hydrogen spillover effect on supported nanoparticle catalysts \cite{Zhang2023} \cite{Karim2017}, and by its propagation to hydrogen driven reaction-diffusion fronts observed at different length scales. \cite{Zeininger2022} \cite{Zeininger2022a} 

The reductive front is structurally propagated until stopped by a too large deviation of the material structure or until locally depleted by the increasing of the reduced area around the pore. This is supported by the earlier observations about the asynchronous and localized behaviour of the surface reduction fronts: only around defected areas such as the feature in \figref{fig:hmr} do these dynamics occur, and the reductive fronts tend to form in similar locations, with similar stopping points between cycles.

\section{Discussion}
The results shown here are purely obtained by analysis of the ESEM images acquired using the dual magnification method. Using only this technique, we were able to obtain a meaningful cross-scale correlation.

Using the dual magnification method, redundant frames are exchanged for frames with an alternate field of view. In a standard experiment, the singular field of view gets images at the set scan rate, whereas with this magnification alternation method, each field of view gets images three times less frequently, particularly suited for long-term experiments to study the deactivation behavior of the catalyst or for systems, in which surface dynamics are slow. 

An alternative method for obtaining multi-scale data would be to simply conduct the same experiment twice. However, compared to the dual magnification method presented here, this has multiple drawbacks. The dual magnification approach simplifies the correlation of information obtained from different length scales, which increases the reliability of the data. Conversely, correlating data from two independent experiments conducted at different magnification can be more difficult. In particular, it is often difficult to reproduce  the observed dynamics exactly with a similar feature in view.

The dual magnification method shown here serves as a demonstration of the power of simple automation of electron microscopy experiments of dynamic systems. Within this field, there are many unexplored avenues, which need to be uncovered to fill blank space in the structure-function landscape of functional materials.

\section{Conclusion}
In summary, we showed that using automation in the electron microscope to explore dynamic systems is a powerful tool to prospectively unleash more realistic structure-function correlations. The described dual magnification algorithm allows for the collection of a larger data diversity, which in turn facilitates structural information under reaction conditions. By comparing simultaneously captured data at multiple magnifications, a phase difference between surface and pore system events in the system was discovered, which would not be possible without the dual magnification method. As such it is a useful tool to enhance our understanding of the working principles of functional materials and demonstrates how different length scales interplay in catalytic processes.

\section{Acknowledgements}
We acknowledge funding from the German Federal Ministry of Education and Research in the framework of the project Cat\-Lab (03EW0015A). This project was in part funded by the Deutsche Forschungsgemein-schaft (DFG, German Research Foundation) - 388390466-TRR 247, subprojects B6 and A4. This work was further supported by the DFG under Germany’s Excellence Strategy – EXC 2089/1 – 390776260. This research was in part funded by the Austrian Science Fund (FWF) [10.55776/F81 and 10.55776/COE5] (SFB TACO, Cluster of Excellence Materials for Energy Conversion and Storage, MECS).

\section{Competing Interests}
The authors declare no competing interests.

\section*{Supporting Information}

\subsection*{SI1: Previous oxidation experiment}
\begin{figure}[H]
    \centering
    \includegraphics[width=1\textwidth]{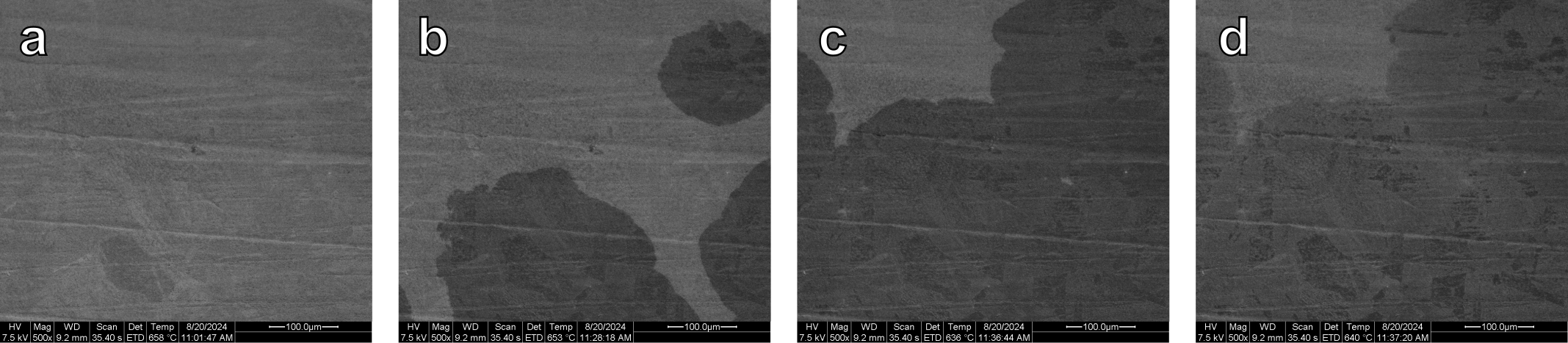}
    \caption*{Figure SI1: Catalyst surface under oxygen darkens when exposed to oxygen.}
    \label{fig:si1}
\end{figure}

In a previous experiment using the same Co catalyst, not in oscillatory conditions, a correlation between the surface intensity visible by the ESEM and the surface state was established. In the figure, the oxidized state is visible in \textbf{Figure SI1 a} under oxygen flow. Shortly before \textbf{Figure SI1 b}, the flow of hydrogen gas was enabled. Following this, the surface quickly started to show the reduced dark areas. During the flow of hydrogen, the reduced area continuously expands, as can be seen in \textbf{Figure SI1 c}. Before the whole surface could be reduced, the flow of hydrogen was disabled again, immediately beginning to re-oxidize the surface (lightening) as seen in \textbf{Figure SI1 d}.

\subsection*{SI2: Full dual magnification video}
In \cite{3.YFJ9MR_2025} a full video of the data that was analyzed in this work can be found. It was taken of a Co foil catalyst surface at 550 \textdegree{}C under 50:1 H$_2$:O$_2$ at high vacuum pressure ($10^{-3}$ Pa) using the FEI Quanta FEG ESEM.

\printbibliography

\end{document}